\documentclass{IEEEcsmag}

\usepackage[colorlinks,urlcolor=blue,linkcolor=blue,citecolor=blue]{hyperref}
\usepackage{orcidlink}
\expandafter\def\expandafter\UrlBreaks\expandafter{\UrlBreaks\do\/\do\*\do\-\do\~\do\'\do\"\do\-}
\usepackage{upmath,color}

\jvol{XX}
\jnum{XX}
\paper{8}
\jmonth{August}
\jname{IEEE Internet Computing Magazine}
\pubyear{2024}

\begin{document}

\sptitle{FEATURE ARTICLE: GENERATIVE AI FOR THE INTERNET OF THINGS}
%\sptitle{INVITED ARTICLE: GENERATIVE AI FOR INTERNET OF THINGS}

\title{The Internet of Things in the Era of Generative
AI: Vision and Challenges}

\author{Xin Wang \orcidlink{0009-0007-6483-9357} and Zhongwei Wan \orcidlink{0000-0002-2790-0290}}
\affil{The Ohio State University, USA, OH 43210}

\author{Arvin Hekmati \orcidlink{0000-0003-0017-9701} and Mingyu Zong \orcidlink{0009-0009-8523-0537}}
\affil{University of Southern California, USA, CA 90089}

\author{Samiul Alam \orcidlink{0000-0002-8458-4642} and Mi Zhang \orcidlink{0000-0001-7002-6757}}
\affil{The Ohio State University, USA, OH 43210}

\author{Bhaskar Krishnamachari \orcidlink{0000-0002-9994-9931}}
\affil{University of Southern California, USA, CA 90089}

\markboth{GENERATIVE AI FOR INTERNET OF THINGS}{GENERATIVE AI FOR INTERNET OF THINGS}

\begin{abstract}
Advancements in generative AI hold immense promise to push the Internet of Things (IoT) to the next level. In this article, we share our vision on the IoT in the era of generative AI. We discuss some of the most important applications of generative AI in IoT-related domains. We also identify some of the most critical challenges and discuss current gaps as well as promising opportunities on enabling generative AI for the IoT. We hope this article can inspire new research on the IoT in the era of generative AI.
\end{abstract}

\maketitle
\section{INTRODUCTION} 
\label{sec:intro}
\vspace{1mm}
\chapteri{T}{oday}, Internet of Things (IoT) such as smartphones, wearables, smart speakers, and household robots have already become an integrated part of our daily lives. 
These devices can sense, communicate, and are empowered by modern artificial intelligence (AI) techniques.

Advancements in Generative AI$^1$ have enabled a new wave of AI revolution. Generative AI refers to AI models that can generate new content in the form of text, images, videos, codes, and many more. While Generative AI is not new, it is only until recently that large-scale generative models exemplified by Large Language Models (LLMs) (e.g., GPT, LLaMA, and Gemini)$^2$ and Multimodal Generative Models (e.g., GPT-4V, DALL-E, and Stable Diffusion)$^3$ have made the breakthrough. Such breakthrough comes from their significantly large model sizes while being pre-trained on massive amounts of data. These characteristics enable Generative AI to generate high-quality data, tackle complex tasks with human-level performance, and exhibit superior generalization ability on new tasks, all of which were not attainable before.

\begin{figure} [t]
    \centering
\includegraphics[width=0.49\textwidth]{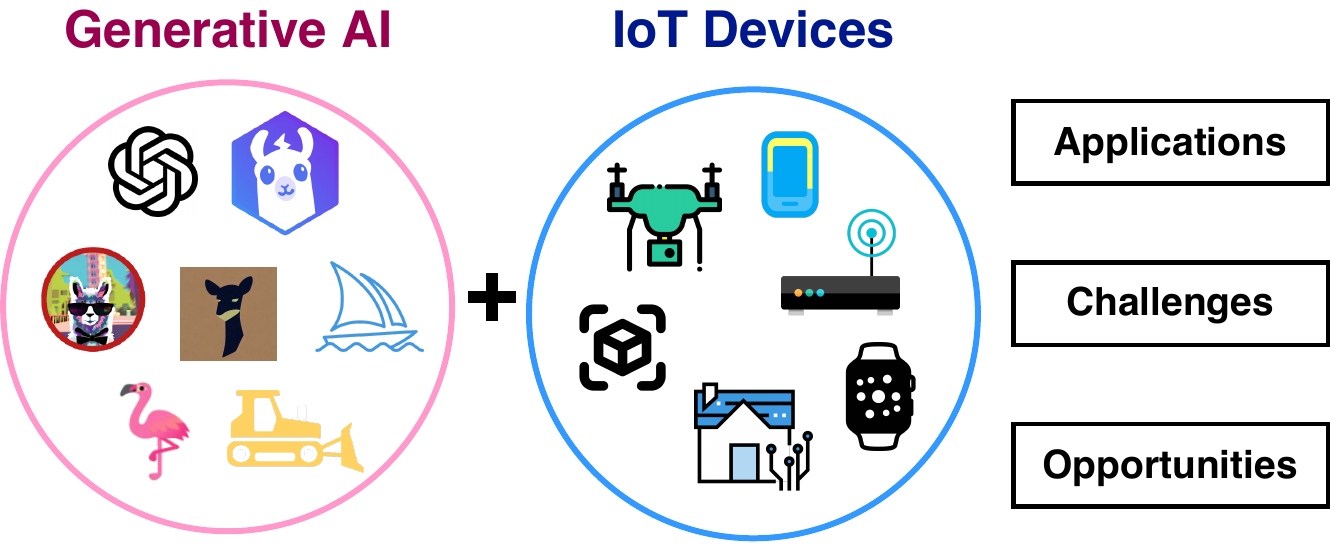}
    \vspace{-0mm}
    \caption{IoT in the era of Generative AI.}
    \label{fig:vision}
    \vspace{-1mm}
\end{figure}

% P3. Opportunities: Briefly talk about the motivation of this paper: 
The implications of the advancements of Generative AI for IoT are profound. The distinctive capabilitie of Generative AI bring pivotal benefits across the entire IoT pipeline, encompassing data generation, data processing, interfacing with IoT devices, and IoT system development and evaluation. These benefits position Generative AI as having substantial potential to revolutionize a wide range of IoT applications such as mobile networks, autonomous driving, metaverse, robotics, healthcare, and cybersecurity. At the same time, turning these applications into reality is, however, not trivial. Innovative techniques are needed to address some of the most formidable challenges so as to realize the full potential of Generative AI for IoT. 

% P5. Overview of this paper.
In this article, we provide our vision and insights on the applications, challenges, and opportunities of what Generative AI brings to IoT (Figure~\ref{fig:vision}). We start by explaining how Generative AI could benefit some of the most important IoT applications. Next, we discuss some of the most critical challenges that serve as impediments to enabling Generative AI for IoT, and share our views on the gaps as well as promising opportunities to address those challenges. We hope this article can act as a catalyst to inspire new research on IoT in the era of Generative AI.

% \input{Section/2_background}
%\input{Section/3_benefits}

% Integration of generative AI into mobile applications and services
\section{APPLICATIONS OF GENERATIVE AI IN IOT-RELATED DOMAINS}
\label{sec:applications}
\vspace{2mm}

\begin{figure} [t]
    \centering
\includegraphics[width=0.34\textwidth]{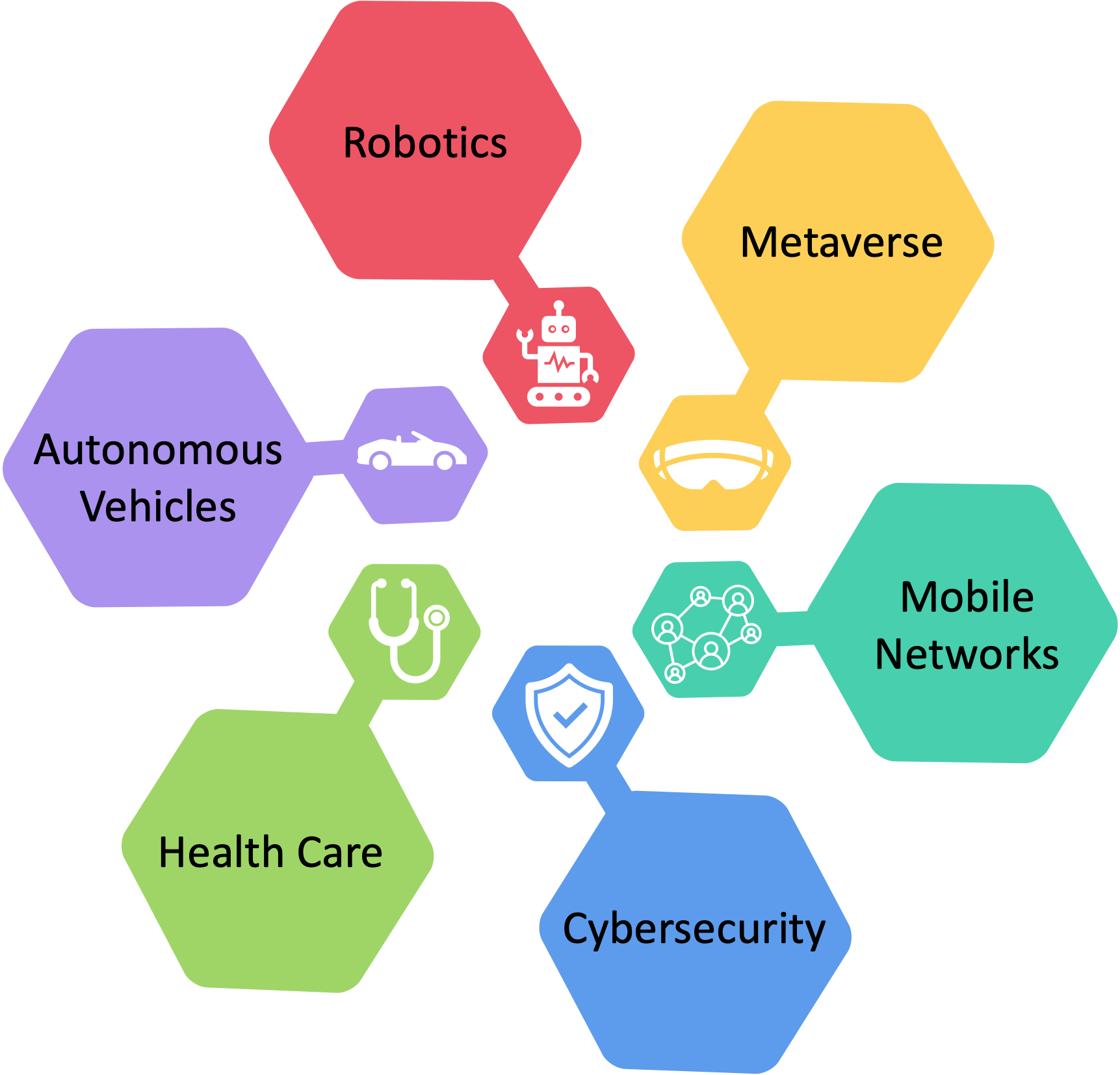}
    \vspace{1mm}
    \caption{Representative application domains of Generative AI for IoT.}
    \label{fig:applications}
    \vspace{-0mm}
\end{figure}

%% Overview:
Leveraging its distinctive capabilities, Generative AI holds the potential to revolutionize numerous critical IoT applications.
% with different types of generated data as shown in Table~\ref{table:application_domains}. 
%While many of these domains have only scratched the surface of what is achievable, others have begun to see profound transformations. 
In this section, we delve into a number of application domains (Figure~\ref{fig:applications}) where Generative AI has already left its mark and others where its potential is just beginning to be recognized.

\subsection{Mobile Networks}
Generative AI has the considerable potential to revolutionize the design and operation of mobile networks$^4$. 
For instance, Generative AI can generate simulations based on historical mobile network data to help network operators foresee potential bottlenecks and adjust resources dynamically. 
% Generative AI 
% %can be employed to optimize network layouts and predict equipment failures before they occur. These models 
% can simulate various network scenarios by generating synthetic data that mirrors real-world conditions, enabling network designers to test and refine network configurations efficiently. 
% %
% Furthermore
Moreover, Generative AI can create highly efficient codecs, resulting in smaller sizes of data
to be exchanged between nodes to increase communication efficiency in mobile networks.

\subsection{Autonomous Vehicles}
The transformational journey of the automobile industry towards autonomous vehicles has been deeply influenced by Generative AI as well$^5$. For example, AI agents empowered by LLMs like Grok can be integrated into vehicle systems such as Tesla Model-3 to enhance user interfaces and improve communication between the vehicle and its occupants, making interactions more responsive and user-friendly. Meanwhile, multimodal generative models are crucial for the development of the advanced driver-assistance systems (ADAS) in autonomous vehicles. ADAS will provide highly accurate  prediction of traffic conditions, pedestrian behavior, and potential hazards, allowing for safer and more reliable vehicle automation. Furthermore, by generating realistic driving scenarios during the testing phase, multimodal generative models will enable manufacturers to refine vehicle responses under various conditions, which significantly accelerates the safe deployment of autonomous vehicles.

\subsection{Metaverse}
Generative AI can significantly enhance the Metaverse by leveraging its ability to visualize and simulate based on multimodal sensor data, generating a vivid and immersive virtual realm$^6$. Moreover, utilizing generative models to adeptly handle data from various sensors and human inputs can construct dynamic, responsive environments that closely mimic the real world or create entirely new, fantastical settings. 
For instance, multimodal generative models such as GPT-4V are capable of generating simulations based on user interactions and environmental changes captured through IoT devices such as smart glasses and head-mounted devices, enabling real-time adjustments and enhancements to the virtual landscape. This capability allows for a seamless and adaptive user experience to make the Metaverse not just a static backdrop but a living, evolving entity.

\begin{figure*}[t]
    \centering
    \includegraphics[width=\textwidth]{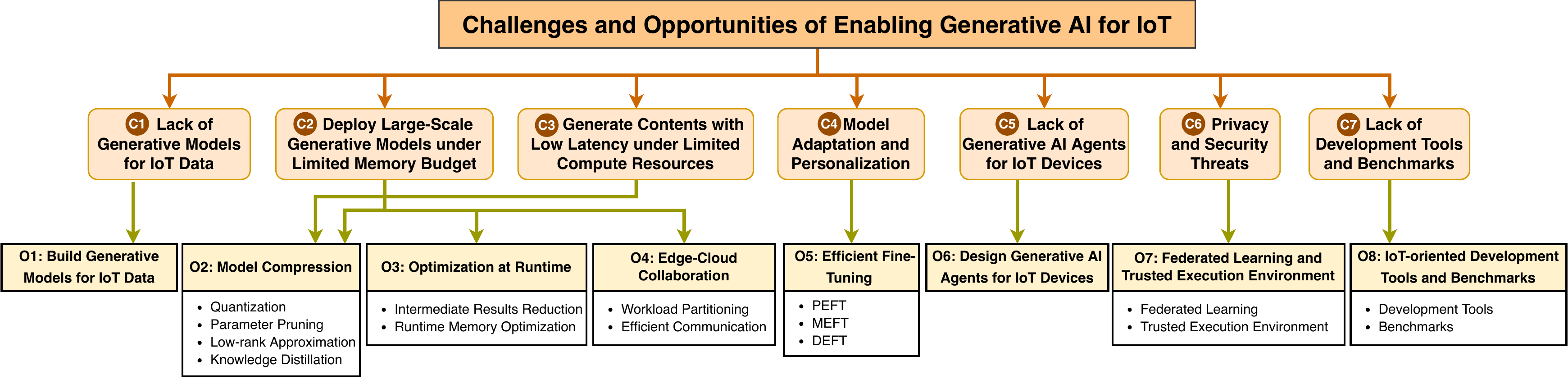}
    \vspace{-3mm}
    \caption{Challenges and opportunities of enabling Generative AI for IoT.}
    \label{fig:challenges}
    \vspace{-0mm}
\end{figure*}

\subsection{Robotics}
In the rapidly evolving field of robotics, the integration of Generative AI has emerged as a cornerstone, significantly enhancing the capabilities and intelligence of robotic systems$^7$. %This integration is further empowered by the use of advanced generative models. 
For example, LLMs can be utilized to enhance the interaction capabilities of robots, enabling them to understand dialogues from users and generate human-friendly responses in real-time. Additionally, multimodal generative models which can effectively process and integrate visual and auditory information, are crucial for robots to better understand their physical environment and context, allowing robots to adapt to new situations with greater autonomy. 
%By generating predictive contents and simulations
In doing so, these powerful generative models help robots learn from their surroundings and make informed decisions, paving the way for more adaptive and intelligent robotic systems in various settings.

\subsection{Health Care}
The healthcare sector, empowered by IoT devices, is experiencing a transformative paradigm shift with the integration of Generative AI to revolutionize patient care$^8$. 
For example, LLMs can be used to automatically process and understand medical literature, and draft patient reports and personalized treatment plans based on patient history.
%process and generate medical texts, automatically generating patient reports,  and even drafting personalized treatment plans based on patient history. 
Meanwhile, multimodal generative models are instrumental in analyzing diverse data modalities collected from IoT devices such as electrocardiogram (ECG), blood pressure monitor, and pulse oximeter; they are also capable of translating those raw sensor data into human-understandable reports$^{9}$. 
These models altogether can integrate text data from electronic health records, numerical data from monitors, and visual data from scans to provide a comprehensive understanding of a patient's health. This integration allows for delivering more accurate diagnoses, preventive care, and tailored treatment regimens, thereby significantly enhancing patient outcomes and operational efficiencies in medical facilities.

\subsection{Cybersecurity}
IoT devices are particularly vulnerable to cyber-attacks due to their widespread deployment and the often minimal built-in security features. Generative AI can be employed to enhance security measures and address their security challenge$^{10}$. For instance, LLMs can analyze and generate security protocols by learning from vast databases of cybersecurity incidents and responses. Similarly, multimodal generative models can integrate and analyze data from multiple sources including network traffic, user behavior, and anomaly detection systems, can predict potential security breaches before they occur. These generative models are trained to identify patterns indicative of cyber threats, enabling proactive security measures and automatic updates to defense strategies, thus protecting the IoT devices against a wide range of cyber threats.

%\subsection{Education}
%To the best of our knowledge, there is no accomplished study centered on the exact convergence of Generative AI, IoT, and Education. Nevertheless, considering Generative AI and IoT's respective strengths, we can envision how they will collaborate and propel the field of education forward. 
%
%Generative AI has exhibited significant potential in the domain of education. Recent studies have validated LLMs' capacity in course content and learning objectives generation \cite{Leiker2023PrototypingTU, Sridhar2023HarnessingLI}. Meanwhile, the research community acknowledges their potential in promoting and delivering personalized learning \cite{Pataranutaporn2021AIgeneratedCF}. It is undeniable that Generative AI models, particularly LLMs, will assume responsibility for creating educational materials. On the other hand, IoT devices will deliver those generated contents across regions. For example, the IoT-based flipped learning platform (IoTFLiP) adopts IoT infrastructure to deliver personalized medical education \cite{Ali2017IoTFLiPIF}. 

%\subsection{Robotics}
%\subsection{Autonomous Driving}
%\subsection{Health Care}
%\subsection{Scientific Writing}
%\subsection{Art Composition}
%\subsection{Personalized Recommendation}
%\subsection{Metaverse and Virtual Reality}

\label{applications}
\vspace{0mm}

\section{CHALLENGES AND OPPORTUNITIES OF ENABLING GENERATIVE AI FOR IOT}
%\section{Opportunities for Training and Deploying Generative AI Models in IoT Systems} 
%\section{Opportunities and Challenges}
\label{sec:challenges}
\vspace{2mm}

Turning these applications into reality is, however, not trivial. We have identified five challenges that act as key barriers to realizing Generative AI for IoT. In this section, we describe these challenges, and share our perspectives on promising opportunities to address these challenges (Figure~\ref{fig:challenges}).

\begin{table*}[h!]
% \centering
 \caption{Representative use cases and data modalities in different IoT application domains.}
%\caption{Summary of generated IoT data with various modalities in different IoT-related application.}
 \vspace{1mm}
\def\arraystretch{1.35}%
\resizebox{1\textwidth}{!}{%
\begin{tabular}{|l|l|l|}
 \hline
\textbf{IoT Application Domain} & \hspace{5mm} \textbf{Representative Use Cases} & \hspace{30mm}\textbf{Data Modalities}                                    \\ \hline
Mobile Networks                 & Network Optimization, Fault Detection & Network Traffic, Network Logs, Multimedia Data    \\ \hline
Autonomous Vehicles             & Navigation, Adversarial Testing       & Lidar, Radar, Point Cloud, GPS Coordinates, Control Signal \\ \hline
Metaverse                       & Virtual Meetings, Virtual Tourism & Gestures, Voices, Facial Expressions, Eye-Tracking Data   \\ \hline
Robotics                        & Path Planning, Human-Robot Interaction    & Images, Audios, Paths, Control Signal, Error Logs    \\ \hline
Health Care                     & Disease Diagnosis, Report Generation     & ECG, Blood Pressure, Steps Taken, Calories Burned, Food Intake   \\ \hline
Cybersecurity                   & Incident Response, Malware Detection  & Authentication Data, Attack Vectors, Breach Reports        \\ \hline
\end{tabular}
}
\label{table:application_domains}
\end{table*}

\vspace{-3mm}
\subsection{Challenges}
\vspace{2mm}
\noindent\textbf{C1: Lack of Generative Models for IoT Data.}
Generative models today are predominantly developed using a few data modalities, including text, images, and videos. However, as summarized in Table~\ref{table:application_domains}, IoT-related applications encompass a much wider range of data modalities, including network traffic data, home-deployed sensor data such as temperature and humidity, as well as healthcare data collected from mobile and wearable devices such as heart rate, steps taken, and calories burned. 
%, which are often overlooked in generative AI research. 
%The lack of generative models for IoT-related data modalities presents a significant gap.
%
%
Therefore, building generative models based on IoT-related data modalities is much needed, yet it remains a very challenging task.
%Building generative models based on IoT-related data modalities is a much-needed but challenging task.
%Exploring the generation based on these diverse IoT modalities presents a promising but challenging topic. 
%Extending generative AI capabilities to encompass IoT data not only enhances the versatility of generative models but also bridges the critical gap between generative AI and the widely deployed IoT ecosystem. %we can unlock new opportunities for innovation in smart environments, industrial automation, healthcare monitoring, and beyond. This expansion not only enhances the versatility of generative models but also bridges the gap between generative AI and the rapidly growing IoT ecosystem.
%, paving the way for more integrated and intelligent systems.

\vspace{2mm}
\noindent\textbf{C2: Deploy Large-Scale Generative Models under Limited Memory Budget.}
Generative models in general contain billions of parameters. Such large model sizes directly translate to their significant demands for memory resources$^{11}$. For example, LLaMA-7B requires about 14GB memory for storing its 7 billion parameters in half-precision floating-point format (FP16). However, IoT devices are known to be memory-constrained. This discrepancy between intensive memory requirements of generative models and limited memory budgets of IoT devices poses a considerable challenge on the deployment of large-scale generative models on IoT devices. %Therefore, how to make these advanced generative models consume less memory to fit in the IoT devices with minimum degradation of their generation ability is urgent to resolve this challenge.
%, including the loading, inferencing, and fine-tuning

\vspace{2mm}
\noindent\textbf{C3: Generate Contents with Low Latency under Limited Compute Resources.}
Many applications of Generative AI in IoT-related domains such as autonomous driving, Metaverse, and robotics are latency-sensitive.
Generative models often involve computation-intensive operations during content generation.
%need to process high-dimensional activation with computational-intensive operations, such as the attention mechanism in LLMs during content generation. 
However, IoT devices are limited in their onboard compute resources, making it challenging to generate contents while meeting application-specific latency requirements. For example, Mixture of Experts (MoE) based LLM ST-l128 requires six seconds to generate only one token on Raspberry Pi 4B$^{12}$, which is too slow for latency-sensitive applications. 
%Therefore, a new resource scheduling and memory management technology is required to enable the fast execution of generative models on resource-constraint IoT devices.
%low latency to enhance user experience.

\vspace{2mm}
%Ensure Robust Generation with Low-quality or Diverse Data on Devices
\noindent\textbf{C4: Model Adaptation and Personalization.} 
As the environments in which IoT devices are deployed evolve, the newly collected data may deviate from prior distributions. Consequently, the post-deployment pre-trained generative models often necessitate fine-tuning on the devices to effectively adapt to this new data. 
Moreover, pre-trained generative models also need to adapt to the local data on the device for personalization to enhance user experience.
%These practical challenges underscore the need for the development of highly efficient fine-tuning techniques to enable on-device fine-tuning for resource-constrained IoT devices. 
%Generative models are pre-trained with high-quality data on server before deployment. However, real-world data can be noisy, incomplete, or out of distribution, making it challenging for pre-trained generative models to generate content with high quality on devices. 

\vspace{2mm}
%\noindent\textbf{C5: Dynamic IoT Task Planning and Flexible Execution.}
\noindent\textbf{C5: Lack of Generative AI Agents for IoT Devices.}
One of the killer applications of Generative AI is AI agents, which are autonomous programs capable of generating new content, making decisions, and performing tasks based on their learned knowledge and user requests.
%. primarily focus on creating data or content, 
%
Currently, most AI agents are designed for cloud platforms and operate as cloud services.
However, due to the privacy concerns associated with personal data collected by IoT devices, along with the high latency of delivering cloud-based services$^{13}$, there is a strong need to develop IoT-based AI agents that can process personal data locally on the devices and deliver a wide range of IoT-oriented services promptly, which is not a trivial task due to the diverse and dynamic nature of IoT ecosystems.
%but they are not inherently designed to integrate with the diverse and dynamic nature of IoT ecosystems. 

\vspace{1mm}
%\noindent\textbf{C4: Guarantee Privacy and Security to Run Generation.} 
\noindent\textbf{C6: Privacy and Security Threats.} 
Data collected by IoT devices such as personal instructions, dialogues, photos and videos often contain privacy-sensitive information that need to be securely stored and processed$^{10}$.
Dealing with privacy and security threats is challenging in the context of generative models. 
%the retrieval augmentation, one of the key but unique procedures during generation, suffers from severe privacy disclosure and cyber attacks. Specifically,
For example, during retrieval-augmented generation (RAG), the query sentences are frequently sent over the network to a remote vector database to find similar samples for augmentation. While being transmitted through the network, the query is vulnerable to leakages or attacks, presenting challenges for protecting the privacy and securing the generative models.
%
%In addition to preserving the privacy of user data captured on IoT devices, ensuring the security of data storage, transmission, and processing for Generative AI applications in the context of IoT is also critical.

\vspace{1mm}
%
%\noindent\textbf{C5: Lack of Software Support and Evaluation for On-devices Generative Models.} 
\noindent\textbf{C7: Lack of Development Tools and Benchmarks.} 
%Currently many frameworks, such as DeepSpeed, vLLM, have been designed to benefit the deployment of generative models from server side. However, the deployment from the device side, such as the mobile phone, is challenging due to the lack of software support. 
%
The implementation of Generative AI for IoT applications presents a wide range of challenges due to the unique characteristics of IoT devices and their deployment environments. To facilitate implementation and widespread adoption of these applications, development tools are essential. 
%as well as LLM-focused development tools such as DeepSpeed and Megatron
%. At the same time,
%,  and ONNX have been developed to deploy models on IoT devices, but unfortunately
However, existing generic development tools such as PyTorch and TensorFlow are not designed for IoT-oriented scenarios; development tools designed for mobile and edge devices such as PyTorch Mobile and TFLite do not provide dedicated supports for large generative models. 
%Moreover, there is a lack of evaluation of existing generative models on devices to test their real performance, speedup after deployment. Therefore, it is still difficult for researchers and practitioners to decide which generative model to use and how to use on IoT devices.
%
Moreover, the advancement of a field cannot be realized without established benchmarks. Unfortunately, there are still no benchmarks that are specifically designed for Generative AI for IoT-oriented applications.

\vspace{-3mm}
\subsection{Opportunities}

\vspace{1mm}
\noindent\textbf{O1: Build Generative Models for IoT Data (Address C1).}
%Extending generative AI capabilities to encompass IoT data not only enhances the versatility of generative models but also bridges the critical gap between generative AI and the widely deployed IoT ecosystem
The lack of generative models for IoT data hinders the development of generative AI-assisted IoT applications. Therefore, there is a significant opportunity for researchers and practitioners to develop new generative models for analyzing or generating various IoT data. 
%Recent advancements have led to the creation of new methods that enable advanced generative models to analyze different types of IoT data. 
%
One promising approach to building  generative models for IoT data is through multimodal LLMs.
%MEIT$^{9}$ is a developed generative model designed to generate medical reports by analyzing ECG signal data. 
For instance, MEIT$^{9}$ is a multimodal LLM designed to understand and analyze raw ECG sensor data, and generate the analysis reports using human-understandable language. 
%Unlike traditional methods that solely rely on user instructions, 
Specifically, MEIT transforms raw ECG sensor data into  tokens and stores them alongside text tokens for report generation. By fine-tuning the MEIT model with a small set of ECG sensor data, it has demonstrated superior performance in generating expert-level ECG reports for heart disease diagnosis.
%from the raw ECG data obtained from IoT devices.

\vspace{1mm}
\noindent\textbf{O2: Model Compression (Address C2, C3).} 
%to reduce their memory usage and computational cost
To address the challenges of model deployment and content generation under limited memory and compute resources on IoT devices, one effective approach is to compress the generative models$^{14}$ before deployment. Conventional compression techniques designed for models that are much smaller than LLMs require retraining after compression. In contrast, many compression techniques for generative models are post-training-based, which avoid retraining after compression given that retraining generative models is expensive.
In general, compression techniques for generative models fall into four categories: quantization, parameter pruning, low-rank approximation, and knowledge distillation. 
Quantization compresses the model by reducing the precision of the weights and/or activations. Nevertheless, even with the most advanced quantization technique, the highest model compression ratio is limited by the smallest bit width.
Parameter pruning compresses the model by eliminating redundant model parameters. Pruning methods can be classified into structured pruning and unstructured pruning. Unstructured pruning has much more pruning flexibility and thus enjoys a lower accuracy drop compared to structured pruning. However, unstructured pruning incurs irregular sparsification, which makes the resulting pruned models difficult to be deployed on IoT devices due to lack of hardware support. 
Low-rank approximation compresses the model by approximating the weight matrix using the product of two or more smaller low-ranking ones with lower dimensions. 
Knowledge distillation (KD) transfers knowledge from a larger model (the teacher) to a smaller one (the student). Unlike the other three compression strategies, it requires training or fine-tuning for knowledge transfer, making it more expensive to apply. 
Lastly, it is worthwhile to note that these four categories of model compression methods are orthogonal to each other, allowing for their combinations to further compress the models.
Existing methods are able to compress LLMs by up to 80\% with minimum accuracy degradation, enabling the deployment of generative models such as LLaMA-7B on IoT devices. As an example, the quantized 4-bit LLaMA-7B can already be deployed in smartphones while only consuming 4GB memory. 

\vspace{1mm}
\noindent\textbf{O3: Optimization at Runtime (Address C2).}
% Another key challenge of enabling Generative AI for IoT is on-device inference. On-device inference is particularly important for latency-sensitive applications such as Metaverse or scenarios where cloud connectivity is not available.
%
Model compression reduces the memory and compute resource demands of generative models before they are deployed on IoT devices. 
%Another opportunity that addresses the challenges of limited memory and compute resources lies at optimization at runtime.
%model deployment and content generation under limited memory and compute resources lies at optimization techniques at runtime.
%Runtime speedup and memory saving is also critical to deploy generative models under limited memory budget and run generation with low latency. Existing methods can be divided into three main types: computation and memory optimization, cross-processor inference, and runtime adaptation.
%
%\textbullet~\emph{{Computation and Memory Optimization:}}
When performing inference at runtime, intermediate results such as activation outputs and Key-Value (KV) cache have to be computed and stored for further processing. 
These intermediate results consume a significant amount of computation and memory resources at runtime. 
Optimizing the runtime memory of KV cache is a unique challenge that conventional ML models do not have.
Thus, techniques that reduce the intermediate results such as evicting unnecessary items from the cache or compressing the cache contents are fruitful optimization opportunities. 
%Techniques such as evicting unnecessary items from the cache or compressing the cache contents are commonly proposed to reduce both memory demand and runtime latency.
%One of the key techniques is to reduce the size of the intermediate results produced by the generative models. For example, the Key-Value Cache, which occupies a significant portion of memory and computation time in LLMs, is a target for optimization. Techniques such as evicting unnecessary items from the cache or compressing the cache contents are commonly proposed to reduce both memory demand and runtime latency.
% When performing on-device inference, intermediate results such as activation outputs and KV Cache have to be computed and stored onboard for further processing. Moreover, the average generation latency for LLaMA-7B on mobile phones is as slow as seven seconds per token. Therefore, reducing the computation and memory footprint of these intermediate results so as to enhance inference efficiency represents a significant challenge for on-device inference. To address this challenge, we envision that one opportunity lies at preprocessing the input states before feeding them into the generative model to reduce the subsequent computation. Another opportunity lies at  I/O optimization. 

%%
Another strategy for runtime optimization is allocating certain model parameters, intermediate results, and computational tasks at different memory hierarchy. Given the limited onboard GPU memory inside IoT devices, leveraging the larger capacity of CPU RAM and disk storage is a promising opportunity. This approach allows for the storage of larger intermediate results and part of model parameters outside GPU memory to enable the execution of LLMs on resource-constrained IoT devices. 
At the same time, a primary challenge of this strategy is the slow data transfer rate between different memory units. Innovative techniques that can reduce the frequency of data transfers or pipeline data transfer with other operations will have great promise to address this challenge.

\vspace{1mm}
\noindent\textbf{O4: Edge-Cloud Collaboration (Address C2).} 
Given the limited memory and compute resources, some IoT devices may not be able to even run the models that are already compressed. In such cases, it is necessary to offload the execution of part or even the whole model to nearby resourceful edge servers or the cloud$^{15}$. 

The key to such edge-cloud collaboration is the design of effective workload partitioning and efficient communication techniques.
%Existing offloading strategy can be separated into two main types: workload partitioning and efficient communication.
%
%\textbullet~\emph{{Workload Partitioning:}}
Workload partitioning 
%refers to the task of partitioning the generative model between the IoT devices and the nearby resourceful edge server or cloud such that different parts of the generative model are executed in a distributed manner. Such task, however, 
is not trivial since IoT devices, edge servers, and cloud have very different compute, memory, and energy resources, and the available network bandwidths can be dynamic. 
%
%Existing techniques can be divided into heuristic-based or learning-based methods. Heuristic-based methods involve the use of predefined rules or experience-driven schemes to partition the workloads. Learning-based methods, on the other hand, are trained on historical workload data to identify patterns and relationships between different tasks and resources to identify optimal workload partitions for new and unseen scenarios. 
%Although both of these two types of methods could achieve good partitioning in some use cases, 
%and the search space 
Due to the NP-Hard nature of the workload partitioning problem, manually identifying the best-performing partition can be practically infeasible, especially for billion-parameter generative models where the number of potential choices of partition points can be extremely large.
%, or when the number of partitions needed scales up and the partitions need to be performed in real time. 
%
One promising opportunity lies at designing a highly-efficient search-based strategy to automatically search for and identify the partition points that optimize the overall performance.
%In such scenarios that are commonly encountered in IoT applications, designing highly-efficient workload partitioning techniques is an important topic for future research.

%
%\textbullet~\emph{{Efficient Communication:}}
Communication between IoT devices and edge servers or the cloud is often conducted through wireless channel in which bandwidth can become the bottleneck.
To ensure a timely exchange of migrated workloads while minimizing bandwidth usage and power consumption caused by wireless transmission, efficient communication is essential. 
%Techniques such as message compression, data sampling, efficient communication protocols, and edge caching have been proposed to optimize communication in resource-constrained scenarios.
At the same time, due to their large model sizes and the potential large amount of data they need to generate, generative models incur a significant burden on communication.
%, especially in scenarios when the wireless bandwidth becomes limited or Generative AI applications are latency-sensitive.
In such cases, we envision that efficient communication techniques are highly demanded. For example, instead of directly transmitting the model or data, techniques that compress the model, embedding vectors, as well as raw data will become extremely useful.

\vspace{1mm}
\noindent\textbf{O5: Efficient Fine-Tuning (Address C4).}
The needs for model adaptation and personalization underscore the importance of developing highly efficient on-device fine-tuning techniques for resource-constrained IoT devices. 
%to enable on-device fine-tuning
%
At a high level, efficient fine-tuning techniques fall into three categories: parameter-efficient fine-tuning (PEFT), memory-efficient fine-tuning (MEFT), and data-efficient fine-tuning (DEFT).
%
%\textbullet~\emph{{Parameter-Efficient Fine-Tuning (PEFT):}}  
PEFT reduces the computational cost of fine-tuning by selecting only a subset of model parameters for tuning. 
%
% PEFT methods can be in general grouped into three categories: addition-based approach, which inserts small neural modules into the generative models as an adapter for updating, or adds some trainable tokens into the input of some layers; the specification-based approach, which only specifies a small number of parameters in the generative models for fine-tuning while keeping the rest frozen; and the reparameterization-based approach, which transforms the updated matrices into a more efficient form, such as the product of the low-rank ones. 
Among PEFT, low-rank adaptation (LoRA) is one of the most widely used methods. Instead of fine-tuning the full parameters of generative models, LoRA freezes the weights and injects trainable rank decomposition matrices into the model. Through fine-tuning these small matrices, the model is able to achieve comparable performance as full-parameter fine-tuning.
%but greatly reduces both the memory demands and time consumption.
%
Although PEFT is able to reduce the computational cost of the fine-tuning process, it can still incur large memory usage.
%on memory-limited IoT devices. 
%
%\textbullet~\emph{{Memory-Efficient Fine-Tuning (MEFT):}} 
% 
Motivated by this limitation, MEFT focuses on reducing fine-tuning memory footprints by conducting model quantization before fine-tuning, utilizing optimizers that require less memory, or combining the gradient calculation and model parameter updating together.
%Although MEFT can address the shortcomings of PEFT to reduce the runtime memory, it may take longer to complete the fine-tuning process, resulting in higher energy consumption. Additionally, the performance of the fine-tuned generative models may be reduced, which largely limits its use on IoT devices.
%\textbullet~\emph{{Data-Efficient Fine-Tuning (DEFT):}} 
Different from PEFT and MEFT, DEFT achieves efficient fine-tuning from a data-centric perspective. By using a small fraction of the data, DEFT can achieve comparable performance to that obtained from fine-tuning with the entire dataset. Another benefit of DEFT is that it can be combined with PEFT or MEFT to further enhance the fine-tuning efficiency. At the same time, most of the existing DEFT methods heavily rely on manual selection of the small set of data for fine-tuning, which can be difficult to accomplish without domain knowledge. Therefore, an automated data selection scheme would benefit more on applying DEFT for specific IoT application domains.
%In summary, fine-tuning on IoT devices is still a challenging research topics but full of opportunities. 
Existing efforts on efficient fine-tuning for IoT devices are able to fine-tune OPT-1.3B using approximately 4GB and 6.5GB of memory on the OPPO Reno6 smartphone$^{16}$. However, most generative models, especially LLMs, contain much more parameters than 1.3 billion.
Fine-tuning large-scale generative models for IoT devices is still a challenging task but full of opportunities.
%and there is still a long way to go for realizing their fine-tuning on IoT devices.

%allow IoT devices to benefit more from DEFT.

\vspace{1mm}
\noindent
\noindent\textbf{O6: Design Generative AI Agents for IoT Devices (Address C5).}
The privacy and latency issues of cloud-based AI agents motivate the design of IoT-based AI agents.
One fundamental capability of AI agents is task planning, which involves breaking down complex tasks into a sequence of simpler steps that could be automatically accomplished.
However, designing AI agents that can perform effective task planning can be challenging given the resource constraints of IoT devices as well as diversified IoT-related tasks including sensing, user interactions, data management and processing, information retrieval, control and activation. 
%In the context of IoT, 
%%To effectively support the deployment of IoT applications, generative models are often required for breaking down complex tasks into some simpler steps that could be automatically accomplished by utilizing interfaces from the specific IoT devices and platforms. This involves not only learning dynamic IoT task planning but also ensuring the accuracy and flexibility of executing these plans sequentially. However, current generative models typically do not possess the capability to manage these complex interactions and requirements, making their integration into IoT applications challenging. 
%
%Building IoT applications based on generative AI agents presents a promising avenue for better integrating generative models into IoT ecosystems.
%Beyond utilizing the individual LLM to generate content required by IoT devices, generative AI agents also leverage LLMs to make a plan of how to utilize the interfaces and tools on IoT devices for better enhancing the generation. However, designing such powerful generative AI agents is a complex task, particularly for deployment in resource-constrained IoT environments. 
%
In such cases, we envision that opportunities lie at developing techniques that allow agents to make highly effective plans for diversified IoT-related tasks and transform the agent’s plans into low-level instructions compatible with various IoT devices, and scheduling onboard resources to ensure the execution of plans in an efficient manner.
%smooth operation of these general agent systems on IoT devices.
%Several critical challenges need to be addressed, including training the LLM agents to enhance their planning abilities for various IoT tasks, transforming the agent’s plans into low-level instructions compatible with various IoT devices and platforms, and efficiently scheduling computational resources to ensure the smooth operation of these general agent systems on IoT devices.

\vspace{1mm}
\noindent\textbf{O7: Federated Learning and Trusted Execution Environment (Address C6).}
%\textbullet~\emph{{Federated Learning:}} 
%
As a privacy-preserving machine learning paradigm, federated learning (FL) emerges as a solution that can improve the quality of the generative models through the personal data while mitigating privacy risks by keeping the data inside the IoT devices$^{17}$.
%
%Federated learning$^{16, 17}$ is also a useful method to protect the data privacy when training the generative models on IoT devices. 
%
While FL has been intensively studied in recent years, most of the proposed techniques have been developed for models with much smaller scales. The emergence of billion-parameter generative models precludes their complete storage within IoT devices due to resource limitations, presenting new challenges in designing FL frameworks that were not previously encountered.
%
%The rise of large generative models drives the need for training them through federated methods. Nevertheless, most current FL techniques are designed for training compact models that can be entirely accommodated within IoT devices. Unfortunately, the scale of large generative models precludes their complete storage within IoT devices due to resource limitations. How to enable federated training for large generative models on IoT devices is a key challenge. 
To address this challenge, we envision that the opportunities lie in exploring partial training (PT)-based approaches where each IoT device trains a smaller sub-model extracted from the large generative model hosted on the cloud server, and this server model is updated by aggregating those trained sub-models. 
For example, 
%Horvath et al. [49] propose FjORD where sub-models are always extracted from a designated part of the large server model. 
Alam et al. introduces FedRolex$^{18}$, which extracts sub-models from the large server model via a rolling window. Such rolling mechanism results in more stable convergence and ensures that the global model is updated uniformly with superior model quality. 
Trusted Execution Environment (TEE) has become a standard technology to enhance the security of IoT devices. TEE provides a secure area within a processor, ensuring that the data and operations executed within its confines are protected from external threats. 
%This is crucial to IoT-related applications \red{such as mobile payment systems or personal data protection, where maintaining the confidentiality and integrity of data is paramount.}
In the context of generative AI, TEE can be leveraged in many innovative ways. 
For example, to secure the user input to the generative models, one can perform tokenization or embedding extraction of the input inside TEE. This ensures the user input is safeguarded from attacks on local devices before being sent to the generative models for inference.

%%
% \red{Apart from FL and TEE, due to the emergence of retrieval-augmented generation in generative AI, some new technologies will also be designed to specifically protect the privacy and security of the users during the retrieval process. For example, due to the limited memory resources on IoT devices, the retrieved examples are often stored in the remote vector database, and both the query and the retrieved vectors will be transmitted through the Internet. In this case, transmitted vectors are often required to be hashed to prevent them from being attacked. We envision that more techniques will be designed in this avenue in the future.}

\vspace{2mm}
\noindent\textbf{O8: IoT-oriented Development Tools and Benchmarks (Address C7).}
%\textbullet~\emph{{Development Tools:}}
%To fill this gap, 
New development tools that support generative models on mobile and edge devices such as llama.cpp and MLC LLM have recently been developed. However, these newly developed tools are still in their infancy. We envision that further refinement on better supporting IoT-related tasks such as runtime resource management, workload offloading, privacy and security enhancement, as well as efficient fine-tuning is much needed and hence holds great promise.
%a promising opportunity. 
%\textbullet~\emph{{Benchmarks:}} 

%
Lastly, although benchmarks such as MMLU, GSM8K, and MMMU are becoming standards to evaluate the performance of generative models for diverse tasks, there is still no benchmark that is specifically designed for generative AI for IoT-related applications. As the development of generative AI for IoT-related applications is advancing rapidly, we envision that a comprehensive and dedicated benchmark that covers a wide range of IoT-oriented data modalities, tasks, platforms, and evaluation metrics such as latency, memory footprint, and energy consumption will become critical and beneficial to the IoT community.

\vspace{-2mm}

\section{CONCLUDING REMARKS}
\label{sec:conclusion}
\vspace{1mm}
%Generative AI has exhibited significant potential to push IoT to the next level. 
Generative AI has shown immense promise in advancing the capabilities of IoT. In this article, we highlighted the key benefits and elaborated on important IoT applications empowered by Generative AI. We also presented the key challenges and opportunities to enable Generative AI for IoT. We hope this article can spark further research in this exciting field.
%on IoT in the era of Generative AI.
\vspace{-2mm}
\section{Acknowledgement}
\vspace{1mm}
This material is based in part upon work supported by National Science Foundation under award NeTS-2312675 and Defense Advanced Research Projects Agency under contract number HR001120C0160. Any views, opinions, and/or findings expressed are those of the authors and should not be interpreted as representing the official views or policies of the funding agency or the US government.

\def\refname{REFERENCES}

\vspace*{-8pt}

\begin{IEEEbiography}{Xin Wang}{\,}  is a Ph.D. student in computer science and engineering at The Ohio State University, Columbus, OH 43210, USA, advised by Prof. Mi Zhang. His research interests include efficient large language models, machine learning systems, and edge AI. Wang received his master’s degree in computer science and engineering from The Ohio State University. Contact him at wang.15980@osu.edu.
\end{IEEEbiography}

\begin{IEEEbiography}{Zhongwei Wan}{\,} is a Ph.D. student in computer science and engineering at The Ohio State University, Columbus, OH 43210, USA, advised by Prof. Mi Zhang. His research interests include efficient large language models, large multimodal models, and their applications. Wan received his master’s degree in control science and engineering from University of Chinese Academy of Sciences, Beijing 100049, China. Contact him at wan.512@osu.edu.
\end{IEEEbiography}

\begin{IEEEbiography}{Arvin Hekmati}{\,} is a Ph.D. student in computer science at the University of Southern California, Los Angeles, CA 90089, USA, advised by Prof. Bhaskar Krishnamachari. His research interests include machine learning, data-driven algorithms,  anomaly detection, and edge computing. Hekmati received his master degree in electrical and computer engineering from McMaster University, ON L8S 4L8, Canada. Contact him at hekmati@usc.edu.
\end{IEEEbiography}

\begin{IEEEbiography}{Mingyu Zong}{\,} is a Ph.D. student in computer science at the University of Southern California, Los Angeles, CA 90089, USA, advised by Prof. Bhaskar Krishnamachari. Her research interests include large language models for the Internet of Things (IoT) and federated learning. Zong received her master degree in spatial data science from University of Southern California. Contact her at mzong@usc.edu.
\end{IEEEbiography}

\begin{IEEEbiography}{Samiul Alam}{\,} is a Ph.D. student in computer science at The Ohio State University, Columbus, OH 43210, USA, advised by Prof. Mi Zhang. His research interests include federated learning and efficient large language models on IoT devices. Alam received his master degree in computer science from Michigan State University, MI 48824, USA. Contact him at alam.140@osu.edu.
\end{IEEEbiography}

\begin{IEEEbiography}{Mi Zhang} {\,} is an associate professor in the Department of Computer Science and Engineering at The Ohio State University, Columbus, OH 43210, USA. He is a senior member of IEEE and ACM. His research interests include Artificial Intelligence of Things (AIoT), machine learning systems, generative AI, mobile computing, and their domain-specific applications. Zhang received his Ph.D. in computer engineering from the University of Southern California, Los Angeles, CA 90089, USA. Contact him at mizhang.1@osu.edu.
\end{IEEEbiography}

\begin{IEEEbiography}{Bhaskar Krishnamachari}{\,} is a professor and Ming Hsieh Faculty Fellow in electrical engineering at the University of Southern California, Los Angeles, CA 90089, USA. He is a Fellow of IEEE. His research interests include the design and analysis of algorithms, protocols, and applications for next-generation wireless networks, the IoT, distributed systems, blockchain technologies, AI and machine learning, and network economics. Krishnamachari received his Ph.D. in electrical engineering from Cornell University, NY 14853, USA. Contact him at bkrishna@usc.edu.
\end{IEEEbiography}

\end{document}